\documentclass[aps,12pt,onecolumn,pre,superscriptaddress]{revtex4-1}
\usepackage{graphicx,wrapfig}
\usepackage{bm}
\usepackage{epstopdf}
\usepackage{amsmath,amsfonts,paralist}
\usepackage{ amssymb }
\usepackage{mathbbol}
\usepackage{setspace}
\usepackage{color}

\usepackage[hmargin=1in,vmargin=1in]{geometry}
\usepackage{setspace}
\usepackage[font=small,labelfont=bf,
   justification=justified,
   format=plain]{caption}

\usepackage{titlesec}
\titleformat{\paragraph}[runin]
{\bfseries\scshape}{\theparagraph}{1em}{}

\newcommand{\captionfonts}{\normalsize}

\makeatletter  
\long\def\@makecaption#1#2{%
  \vskip\abovecaptionskip
  \sbox\@tempboxa{{\captionfonts #1: #2}}%
  \ifdim \wd\@tempboxa >\hsize
    {\captionfonts #1: #2\par}
  \else
    \hbox to\hsize{\hfil\box\@tempboxa\hfil}%
  \fi
  \vskip\belowcaptionskip}
\makeatother

\usepackage[right]{lineno}

   \let\th=\theta

\let\citep = \cite

\newcommand{\OM}[1]{{\color[rgb]{0.0,0.0,0.0}#1}}

\begin{document}

\title{Predicting synchronous firing of large neural populations\\
from sequential recordings}

\author{Oleksandr Sorochynskyi}
\affiliation{Sorbonne Universit\'e, INSERM, CNRS, Institut de la Vision, 17 rue Moreau, 75012 Paris, France.}
\author{St\'{e}phane Deny}
\affiliation{Neural Dynamics and Computation Lab, Stanford University, California}
\author{Olivier Marre}
\affiliation{Sorbonne Universit\'e, INSERM, CNRS, Institut de la Vision, 17 rue Moreau, 75012 Paris, France.}
\affiliation{Equal contribution}
\author{Ulisse Ferrari}
\affiliation{Sorbonne Universit\'e, INSERM, CNRS, Institut de la Vision, 17 rue Moreau, 75012 Paris, France.}
\affiliation{Equal contribution}
\thanks{Correspondence should be sent to \url{ulisse.ferrari@gmail.com}.}

\begin{abstract}
A major goal in neuroscience is to understand how populations of neurons code for stimuli or actions. 
While the number of neurons that can be recorded simultaneously is increasing at a fast pace, in most cases these recordings cannot access a complete population: 
some neurons that carry relevant information remain unrecorded. 
In particular, it is hard to simultaneously record all the neurons of the same type in a given area. 
Recent progress have made possible to profile each recorded neuron in a given area thanks to genetic and physiological tools, and to pool together recordings from neurons of the same type across different experimental sessions. 
However, it is unclear how to infer the activity of a full population of neurons of the same type from these sequential recordings. 
Neural networks exhibit collective behaviour, e.g. noise correlations and synchronous activity, that are not directly captured by a conditionally-independent model that would just put together the spike trains from sequential recordings. 
Here we show that we can infer the activity of a full population of retina ganglion cells from sequential recordings, using a novel method based on copula distributions and maximum entropy modeling. 
From just the spiking response of each ganglion cell to a repeated stimulus, and a few pairwise recordings, we could predict the noise correlations using copulas, and then the full activity of a large population of ganglion cells of the same type using maximum entropy modeling.
Remarkably, we could generalize to predict the population responses to different stimuli and even to different experiments. 
We could therefore use our method to construct a very large population merging cells' responses from different experiments.
We predicted synchronous activity accurately and showed it grew substantially with the number of neurons. 
This approach is a promising way to infer population activity from sequential recordings in sensory areas.
\end{abstract}

\maketitle
\clearpage

\section*{Introduction}
A major goal of neuroscience is to understand how populations of neurons process sensory stimuli. This understanding is limited because, among other reasons, accessing the activity of all neurons of a sensory structure is very challenging. Most techniques only give access to a small fraction of neurons \cite{Buzaki04,Packer14} (but see \cite{Marre12,Turaga12}), leaving as hidden variables many neurons that may play a role in information processing but are not recorded. 

To overcome this issue, an emerging, `divide and conquer' approach {is to first classify} the neurons in a given area into different cell types, where neurons of the same type are supposed to be functionally identical. {Then, in a second step}, one can characterize the neuronal function of each cell type, to eventually predict {how populations composed of all the neurons of the same type will respond to sensory stimuli.} 

There has been tremendous progress recently in achieving the first step of this approach. Several studies have shown that it is possible to cluster cells in different homogeneous types \cite{Vlasits19}.
This can be done using either the responses of each cell to several standard stimuli \cite{Farrow11,Baden16,Spampinato19}, 
{ or using genetic tools \cite{Economo18,Kim18}.}
These methods have proven successful in isolating most cell types in the retina \cite{Silveira11,Baden16,Shekhar16} and there are {several ongoing studies trying to apply these approaches} in the cortex \cite{Jiang15}.

For the second step, many studies have tried to model and predict how neurons {of a single type} respond to complex stimuli. This strategy has been applied in the retina \cite{Chichilnisky01,Pillow08,McFarland13,Mcintosh16,Deny17} and in many low-level areas \cite{Aertsen80,Sahani03,David04}. A complementary strategy is to use mouse lines expressing GFP in specific cell types in order to record sequentially (i.e. repeatedly across different experiments) from cells that are functionally identical. This has been performed in the retina \cite{Munch09,Farrow11} and enables one to present as many stimuli as desired to cells belonging to the same type. It is thus possible to gather a lot of information about how single neurons of a well-defined cell type will respond to many different sensory stimuli, using sequential recordings of neurons of the same type taken from different experiments. 

Extensive characterization of single cell responses to sensory stimuli is thus possible. The next challenge is to infer how the entire ensemble of neurons of a single type responds together to stimuli. Ideally, one would like to record from all the neurons of a given type, but this is rarely possible. 

One possible strategy is to use these sequential recordings from cells of the same type to reconstruct how the entire population will respond. 
However, reconstructing the activity of a full population from sequential recordings cannot be done by simply pooling the responses to a given stimulus from many sequential recordings. 
{In many cases, pairs of neurons are correlated due to shared noise (noise correlation), which might significantly reshape the neurons' activity \cite{Schneidman03}, and play an important role in information encoding and transmission \cite{Trong08,Pillow08,Shlens08,Greschner11,Lyamzin15,Pitkow12,Franke16}}. 
{Because these noise correlations cannot be predicted from sequential recordings, a model is needed to predict them and therefore to infer the activity of a full population of neurons of the same type. }

Previous works have tried to model and predict the activity of ensembles of neurons in the retina 
{\cite{Schneidman06,Pillow08,Shimazaki12,Granot-Atedgi13,Ferrari18b}}.
However, they were fitted and tested on ensemble of neurons recorded simultaneously, and it is unclear if they can generalize to predict correlated activity across experiments, a critical feature {necessary} to reconstruct activity from sequential recordings. 
Overall firing rates will vary between experiments, and this variation will make these models unlikely to predict noise correlations across experiments. 

Here we address this issue and propose a method to infer the activity of an entire population of neurons of the same type from sequential recordings \OM{in the retina}. 
Our method assumes that we have access to many single cell recordings gathered from different experiments of neurons of the same type, where the same stimulus has been displayed, and additionally to a few recordings of pairs of neurons of the same type. 
We used these data to reconstruct the activity of a large population of neurons, with a shared noise consistent with the paired recordings. 

We applied this method in the rat retina, where the activity of many neurons of the same type can be recorded through repetitions of multi-electrode array experiments \cite{Marre12}, so that the method can be validated.
We first show that a copula-based analysis \cite{Trivedi07,Berkes09,Onken09,Safaai18} of synchronous activity allows a simple description of noise correlations, that is invariant across stimulus identities and experimental preparations.
This description depends only on the individual activities of pairs of cells and on their physical distance.
This result allowed us constructing a model based on copulas and to predict -across experiments- the extent of pairwise noise correlations from sequential recordings of neurons of the same type.
From this estimation of pairwise correlations we then used a time-dependent maximum entropy model \cite{Shimazaki12,Granot-Atedgi13,Ferrari18b} to infer the activity of the full population of neurons of the same type. 
We show that this method is accurate and reproduces several features of the recorded population activity. 
We then applied our method to infer the activity of a large population of neurons of the same type, beyond what can be currently recorded experimentally. 
Thanks to our inference method, we could estimate the extent of synchronous firing in such a large population, and show that it grows significantly for large populations.

{As soon as sequential recordings of many cells of the same type will be available also in cortical systems, it will be possible to apply our method to reconstruct the activity of large populations even beyond the retina.}

\clearpage

\section*{Results}
\subsection*{Overview of the inference method}~

The purpose of our method is to reconstruct the activity of a population of neurons of the same type from their individual responses to a same stimulus. 
Part of this population activity is directly accessible from sequential recordings, but another part needs to be predicted. For example, if we have recorded sequentially two neurons responding to the same stimulus, a naive solution is to pool together their responses as if they had been recorded at the same time. 
If the noise present in these responses is independent between the two neurons, this is indeed equivalent to record them together. However, in many cases, the noise between different neurons is correlated. In that case, pooled sequential recordings are not equivalent to simultaneous recordings \cite{Schneidman03}, and the difference is what is usually termed the noise correlation between these two neurons. 

Our method aims at inferring these noise correlations from parsimonious pairwise recordings {of a few cells}, and use them to predict how noise will be correlated across an entire population of {hundreds of} neurons. From this we can reconstruct how a large population of neurons would respond if they were recorded simultaneously, based on sequential recordings. 

Our method is divided in three steps.
First, we infer the parameter of a model based on copulas \cite{Trivedi07,Berkes09,Onken09} from simultaneous recordings of few cells.
Second, we used the inferred model to predict noise correlations between pairs of neurons from sequential recordings, using only information on the distance between the recorded cells.
Our method allows predicting noise correlations for the same pair of neurons responding to a different type of stimulus, and can generalize to predict noise correlations for another pair of neurons of the same type recorded in a different experiment. 
Third, we use a time-dependent maximum entropy model \cite{Shimazaki12,Granot-Atedgi13,Ferrari18b} to generalize from pairs of neurons to a full population.
This step does not require any additional empirical informationwith respect to the second step.
Note that simultaneous recordings are necessary only for model inference (and validation), yet sequential recording are sufficient for making prediction.

Here we applied this method to cells of the same type in the rat retina.
These data allowed us testing if our reconstruction of the population activity is accurate.
Finally, we used our method to infer the synchronous activity of a large neuronal population, much larger than what is nowadays experimentally accessible.

\subsection*{Strong noise correlations between nearby OFF retinal ganglion cells.}

\begin{figure}[t!]
\centering
\includegraphics[width=\textwidth]{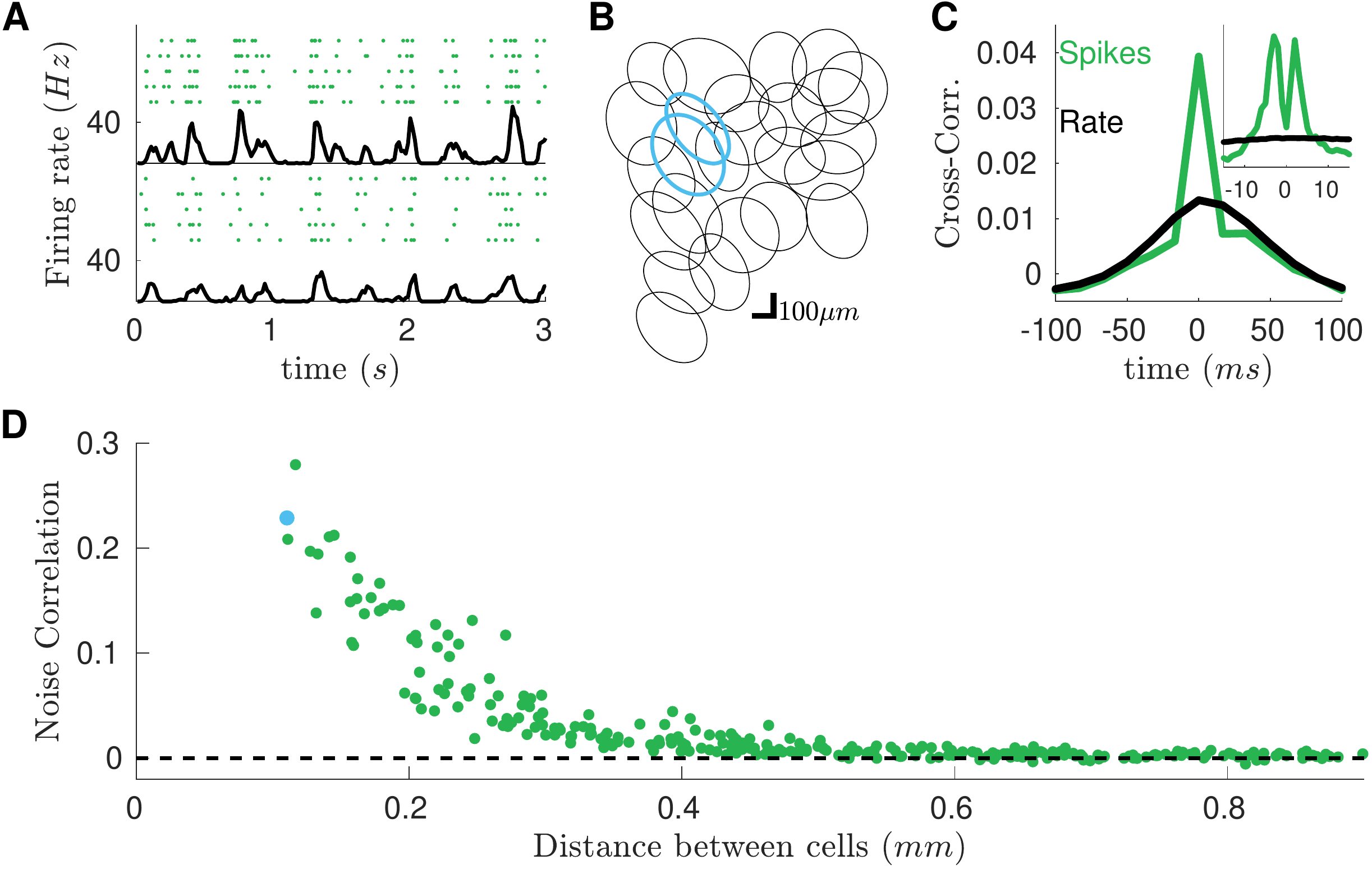}
\caption{
\textbf{High noise correlation between nearby RGCs subject to checkerboard stimulation.}
{\bf A}) Raster plots of two example  cells in response to checkerboard stimulation. Each line corresponds to a repetition of the same visual stimulation.
Black line: averaged firing rate of the cell.
{\bf B}) Receptive field mosaic of the recorded OFF cell population. Cyan receptive fields refer to the cells showed in panel A.
{\bf C}) Cross-correlation for the two cyan cells (green, $dt=17ms$), superimposed to the cross-correlation of their firing rates (black).
Inset: Cross-correlation at finer time scales ($dt=1ms$).
{\bf D}) \textit{Zero-lag} ($dt=17ms$) noise-correlation plotted against the distant between cells. 
}
\label{fig_1}
\end{figure}

We recorded rat retinal ganglion cells (RGCs) in response to different visual stimuli. 
We used a previously described method \cite{Deny17} to divide them in different types. Briefly, we clustered their responses to a full field flicker and isolated a single type of OFF-alpha ganglion cells. 
All these cells responded reliably to a checkerboard stimulus (Fig.~\ref{fig_1}A). 
The cell responses to random checkerboard have been used to estimate their receptive fields and to find their location.
The receptive fields of these cells tiled regularly the visual field (mosaic in Fig.~\ref{fig_1}B).

{To estimate noise correlation between pairs of cells, we computed the cross-correlation of their spike count and the cross-correlation of their firing rate (the mean over stimulus repetitions of the spike count, respectively green and black lines in Fig.~\ref{fig_1}C inset)}.
At short time-scales, spike-count correlation is larger than that of firing rates , but only for nearby pair of cells (see Fig.~\ref{fig_1}C). 
We term noise-correlation the difference between the zero-lag cross-correlation of spike counts and firing rate (see Methods).
We observed a similar behavior for all visual stimulations and experiments.

In the following we used these data to test our method. 
We first used copulas to predict the noise correlations between pairs of cells. 
We then used maximum entropy model to reconstruct the activity of a large population of ganglion cells from sequential recordings. 

\subsection*{Copula model predicts pairwise response from sequential recordings.}

\begin{figure}[t!]
\centering
\includegraphics[width=1.0\textwidth]{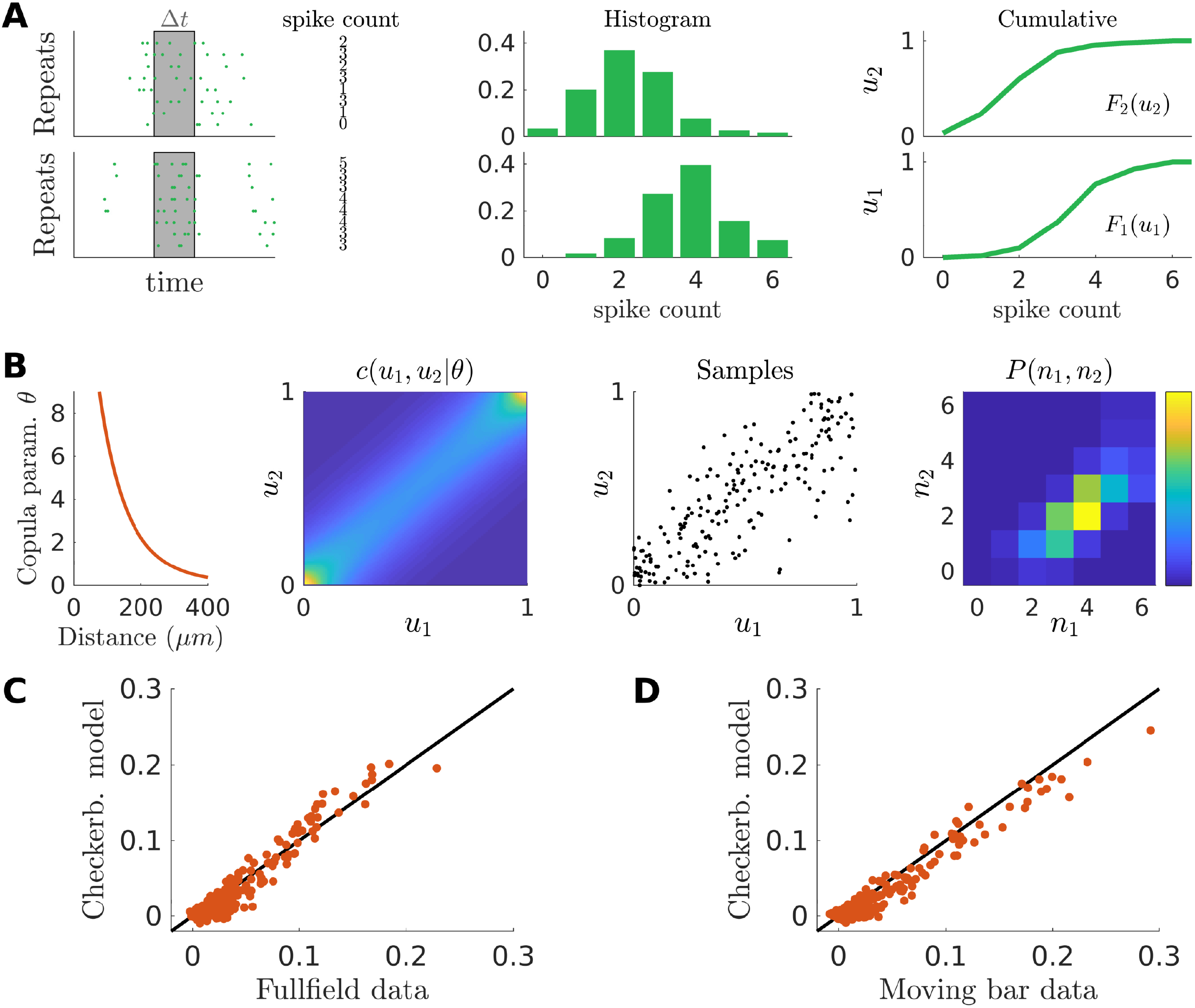}
\caption{
\textbf{Copula model predicts noise correlations across stimulus ensembles.}
{
{\bf A}) Illustration of our copula model: first step. Spike counts are estimated across stimulus repetition for a given pair of cells, in a given time-bin. 
Empirical histogram of the spike counts are then estimated and later used to compute the empirical cumulative distribution function.
{\bf B}) Second step. A parametric function allows to estimate the copula parameter $\theta$ as a function of the cells' distance. A copula distribution then accounts for the mutual dependency of two random variables.}
From it we draw many samples of pairs of real numbers $(u_1,u_2) \in [0,1]^2$ with uniform marginals.
Inverse of the cumulative distribution functions transform these samples into pairs of positive integer numbers, whose distribution matches the empirical spike count marginals by construction and accounts for their mutual dependency.
{\bf C}) Scatterplot of the empirical and model predicted noise correlations fro the response to full-field stimulation.
{\bf D}) As {\bf C} but for the moving-bar stimulation.
In both case the model has been inferred from the response to checkerboard stimulation.
}
\label{figCopPred}
\end{figure}

A copula is a method to build pairwise probability distributions from pairs of single-variable distributions (see Fig.~\ref{figCopPred} and  Methods).
We used this approach to build the joint spike count distribution of pairs of neurons, that, if marginalized, reproduces the empirical single neuron distributions.
For each time-bin, for each recorded neuron, we first estimated the distribution of spike count from its response to stimulus repetitions, and from this we obtained its cumulative distribution function.
Next, for each pair of neurons, we fitted one copula distribution to the collection of joint cumulative functions of activity.
We then drew samples from the inferred copula distribution, i.e. pairs of real numbers between $0$ and $1$ ($(u_1,u_2) \in [0,1]^2$) with uniform marginal distribution. 
Finally, we used the inverse the cumulative distribution functions to transform these samples into pair of integers, which followed the predicted joint spike-count distribution.

A copula is characterized by a parameter that tunes the interaction strength of the two variables, and that can be inferred from data.
We found that this copula parameter only depended on the distance between the two cells, and can be fitted using a  function with just three parameters ($\theta = \exp( a+bx+cx^2)$, where $x$ is the distance between the two cells, see Fig.~\ref{figCopPred}B, suppl. sect.~\ref{suppl:inference} and Fig.~\ref{figCopula}).
We could thus describe the joint activity between {all} pair of neurons using copulas characterized by only three parameters across the entire population of cells.
Once applied on the same response to checkerboard stimulation used for training, our model predicted noise correlations with high accuracy (Pearson's $\rho=0.99$, $n=300$ pairs, {larger than $\rho=0.92$ obtained by fitting directly noise correlations, see supplementary sect.~\ref{suppl:nullModel}}).

We have built a model with only 3 parameters, that can predict the noise correlation between any pair of neurons from the activity of single cells. 
We then tested if this model can generalize and predict noise correlations measured in response to different stimuli. 
We first inferred the copula model from the response to checkerboard stimulation (that of Fig.~\ref{fig_1}).
Then, from the response to repetitions of another type of stimulus, we estimated the spike-count distributions of each neuron in each time-bin.
Finally, we used our copula model to first build the joint distribution, and then compute the mean noise-correlations of each neuron pair.
We applied this strategy to the RGCs' response to full-field and moving-bar stimuli, see Fig.~\ref{figCopPred}.
In both cases, the copula model was able to reproduce the empirical estimates of noise-correlations with high accuracy {(Pearson's $\rho=0.96$ for full-field and $\rho=0.97$ for moving-bar, $n=300$ pairs)}.

\begin{figure}[t!]
\centering
\includegraphics[width=1.0\textwidth]{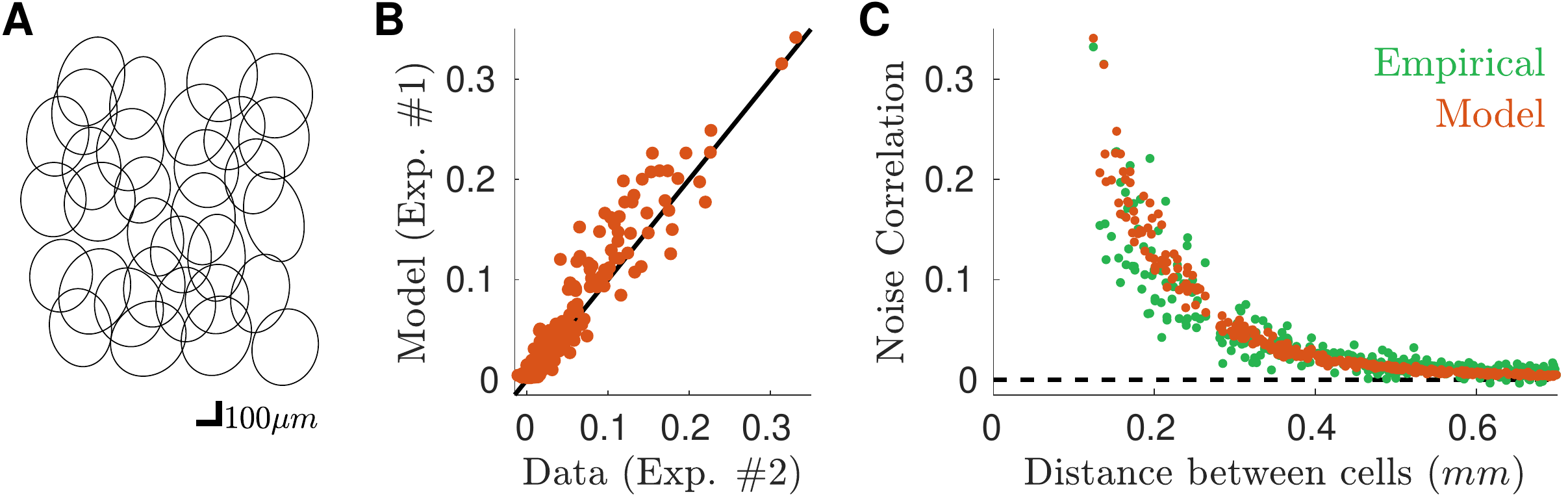}
\caption{
\textbf{Copula model predicts noise correlations across experimental preparations.}
{\bf A}) Receptive field mosaic of the recorded OFF cell population for a new dataset, different from the one used for training the model. 
{\bf B}) Scatterplot of the empirical and model predicted noise correlations from the response to checkerboard stimulation.
{\bf C}) Behavior of the empirical and model predicted noise correlations plotted against the distance between cells. 
Data from a second dataset (\#2), different from the one used for training the model (\#1).
}
\label{figCopExp}
\end{figure}

To further demonstrate the robustness of our method we tested if our copula model could predict noise correlations in a different RGC population of the same type, recorded in a different experimental preparation.
We used the model inferred from the data of the first experiment to predict noise correlations between the same type of RGC, but recorded during a second experiment.
Using only single cell responses, our model predictions were accurate {(Fig.~\ref{figCopExp}B, Pearson's $\rho=0.96$, $n=496$ pairs)}, and accounted for how noise correlations decrease with distance (Fig.~\ref{figCopExp}C).
The functional dependence of the copula parameter with respect to inter-cell distances is thus robust across experiments, and hence corresponds to a general property of OFF-Alpha cells in the rat retina.
We obtained similar results for all the 8 testing experiments (averaged Pearson's $\rho = 0.949 \pm 0.017$, for a total of $n=1632$ pairs).

Note that in order to predict noise correlations (Fig.~\ref{figCopExp}) our model never accessed to the simultaneous recordings, but only to the collection of single neuron responses. 
We could have thus predicted the noise correlation between pairs of neurons in these new experiments using only the sequential recording of each neuron.
Our approach thus allows predicting pairwise noise correlations across experiments without requiring simultaneous recordings.

\subsection*{Time-dependent maximum entropy model reconstructs the activity of large population from the copula's pairwise predictions.}

Our copula model predicted pairwise synchronous firing. To reconstruct the activity of a large population of neurons from single cell recordings, we then used a  \textit{time-dependent} Maximum Entropy population model. 

Standard Maximum Entropy models \cite{Schneidman06} aim at predicting the probability of any spike pattern from the mean firing rate of each neuron and the correlations between each pair of cells. 
Here we use a recent generalization of this approach \cite{Granot-Atedgi13,Ferrari18b} that takes into account a time-varying firing rate. 
This approach built a collection of pairwise Maximum Entropy models (one for each time-bin), which share the same couplings, but with OM{different external inputs (fields)} 
for each cell and each time bin \cite{Granot-Atedgi13,Ferrari18b} (see Methods).
\textit{Time-dependent} Maximum Entropy modelling thus disentangles intrinsic interaction, due to network effects, from extrinsic correlations, due to common inputs \cite{Ferrari18b}.

\begin{figure}[t!]
\centering
\includegraphics[width=1.0\textwidth]{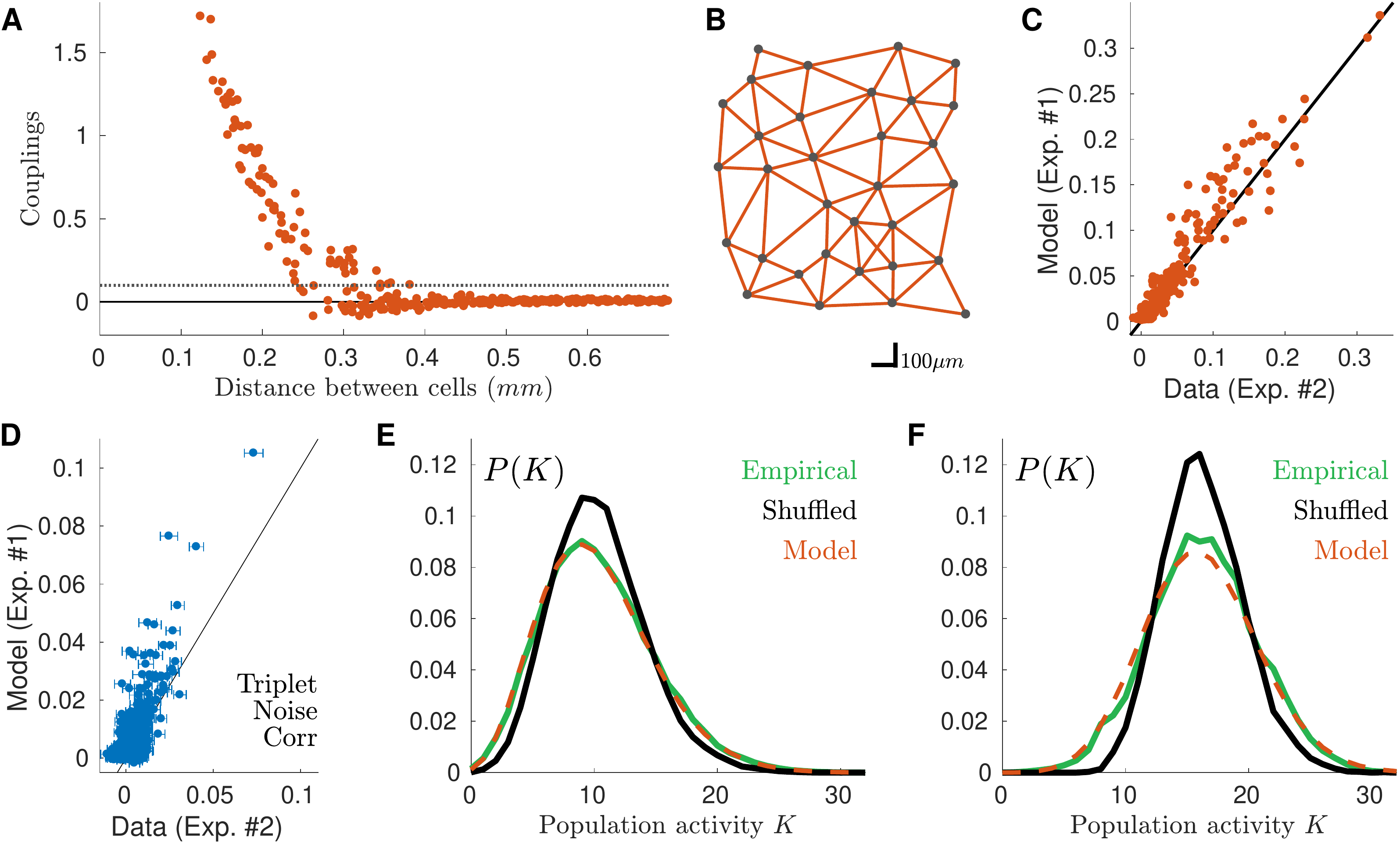}
\caption{
\textbf{Time-dependent Maximum Entropy model predicts population synchronous firing across experimental preparations.}
{\bf A}) Behavior of the inferred couplings with the distance between cells. Grey dotted line threshold for identifying stong couplings (see panel \textbf{B}).
{\bf B}) Position of the cells on the retinal surface, and stong couplings linking them. In the model, only nearest neighbor cells are directly interacting
{\bf C}) Noise correlation prediction of the model against empirical value. As expected by construction, the model reproduces the copula estimations, and hence the empirical values (see Fig.~\ref{figCopExp}B).
{\bf D}) Model prediction of triplet noise correlations against empirical value (Pearson correlation $\simeq$ 0.6, $n=4960$). 
{\bf E}) Empirical, shuffled (cond. independent) and model distributions of the population activity averaged over all time-bins.
{\bf F}) Same as {\bf E} but for the $5\%$ of time-bins with the highest mean population firing rate.
Data from a second dataset (\#2), different from the one used for training the model (\#1).
}
\label{figPK}
\end{figure}

We inferred the \textit{time-dependent} Maximum Entropy population model from the activity of single neurons and the pairwise correlations predicted by the copula model.
The inferred couplings are large only between nearby cells, see Fig.~\ref{figPK}A, and the model reconstructs a ``nearest-neighbor'' interaction network, see Fig.~\ref{figPK}B.
As expected the model reproduced well the pairwise noise correlation (Fig.~\ref{figPK}C, Pearson's $\rho=0.96$, $n=496$ pairs), which were already finely predicted by the copula model (Fig.~\ref{figCopExp}B).
Remarkably, Fig.~\ref{figPK}D shows that the model also accounts for the triplet noise correlation (see Methods) (Pearson's $\rho = 0.6$, $n=4960$ triplets).
In order to show that our model captures the synchronous behavior of the neuronal population, we compute the probability distribution of the population rate, i.e. the total number of spikes emitted by the entire population in a given time-bin.
We compared this distribution with the ``shuffled'' distribution, which destroys noise correlations, and is equivalent to the prediction made by a conditionally-independent model (i.e. a model that would assume there are no noise correlations). 
The distribution computed after shuffling the data overestimated the probability of number of spikes close to the average population rate, and underestimated the occurrence of transients of very large or very low activity (see Fig.~\ref{figPK}E and F). 
{Remarkably, our model captured the empirical behavior of the population rate averaged over the whole recording (see Fig.~\ref{figPK}E). 
It also performed well when focusing only on highly active time-bins (Fig.~\ref{figPK}F).}

\subsection*{High synchrony in a large population of ganglion cells reconstructed from multiple experiments}~

\begin{figure}[t!]
\centering
\includegraphics[width=1.0\textwidth]{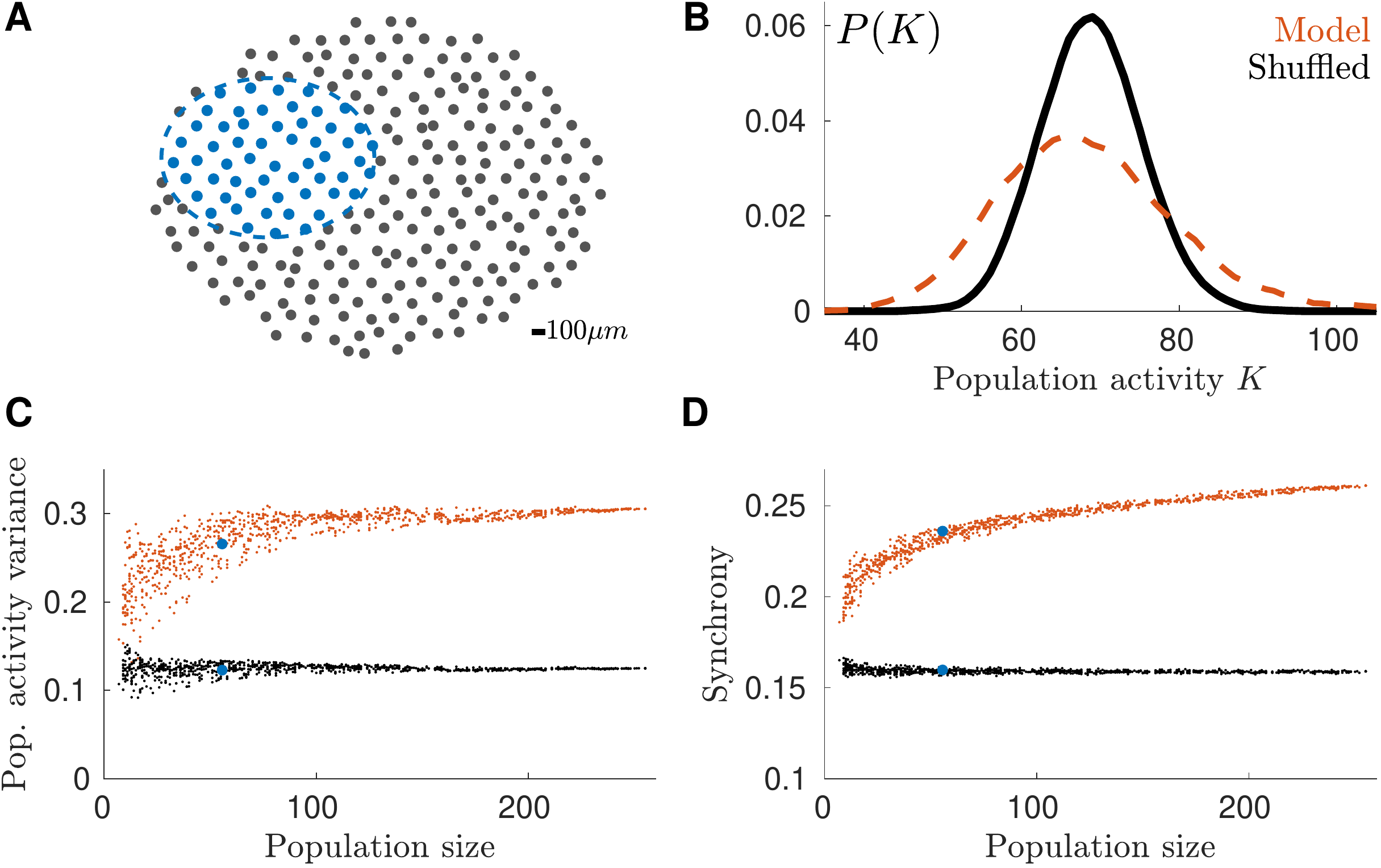}
\caption{
\textbf{Large population model.}
{\bf A}) Synthetic mosaic of $n=256$ cells. Blue: example of sub-population.
{\bf B}) Probability distribution of the population activity for the time-bin with the highest firing rate. 
{\bf C}) Average over time of the population activity variance for many sub-populations.
{\bf D}) Behavior of Synchrony with the number of cell in the sub-population.
}
\label{figSynth}
\end{figure}
Thanks to our model, we could reconstruct the activity of a large population of neurons using only single cell activity. 
Since the model can generalize across experiments, it means that the activity of the different single cells can be taken from different experiments. 
Our method only needs a few pairs of neurons recorded simultaneously to fit the three parameters of the model. 
In the following we illustrate how this model can be used to reconstruct the activity from a very large population of neurons, bigger than what could be recorded experimentally. We illustrate how this inference of the activity of a large population of cells can be useful by measuring synchronous activity over increasing number of cells. 

We collected the (marginal) response of many cells recorded during multiple experiments, and constructed a large population of $n=256$ cells.
We first built a synthetic lattice representing the positions of the cells reproducing the empirical statistics of inter-cells distances (Fig.~\ref{figSynth}A and Methods).
Then, we associated to each lattice position the response to the checkerboard stimulus of a randomly chosen cell among the ones recorded in all experiments (excluding those from the experiment used to learn the model parameters).
Finally we applied our two-step approach to predict how this large population of cell would have responded to a checkerboard stimulus. 

At first, we looked at the population rate as previously defined. 
In comparison with shuffled data, our model predicted a more frequent occurrence of transients with either very high or very low population activity (Fig.~\ref{figSynth}B).
This is a signature that synchronous activity extends up to large populations.
To study how correlated firing grows with the population size, we sub-sampled the synthetic model (Fig.~\ref{figSynth}A) and computed  the variance of the population rate.
The variance predicted by the model grows with the number of neurons (Fig.~\ref{figSynth}C) and saturates at $\sim 200$ neurons ( population activity variance is $0.24 \pm 0.02$ at $N \simeq 30$ cells and $0.299 \pm 0.002$ at $N \simeq 200$ cells , mean $\pm$ s.d.), much larger than the typical number of cells recorded in an experimental session.
On the contrary, in the reshuffled control, the variance is smaller and roughly constant when the number of neurons increases.
Next we estimated the synchrony in the population as the probability of observing a transient with large population activity (see Methods).
This synchrony grew fast for small number of cells {and stopped increasing at $\sim200$ cells} ( synchrony is $0.221 \pm 0.004$ at $N \simeq 30$ cells and $0.257 \pm 0.001$ at $N \simeq 200$ cells , mean $\pm$ s.d., Fig.~\ref{figSynth}D) .

\clearpage
\section*{Discussion}
We have shown that our new method allows for reconstructing the population activity of large populations of neurons of the same type in the retina, based on sequential recordings and a few pairwise recordings.
We have first developed a model to predict noise correlations across experiments. Thanks to this, once the model parameters are learned with paired recordings, we can take any pair of cells taken from new experiments, and predict the noise correlations from the activity of each single cell. 
Thanks to this prediction we could then reconstruct the activity of large populations of neurons of a single type at a scale beyond what can be recorded experimentally. 
Using this method, we have shown that synchrony in the population of OFF-alpha ganglion cells grows with the number of cells, and become large for large populations. Understanding how the collective behaviour of neural ensembles scales with the number of neurons is a crucial issue, and our tool is a key method for this purpose, because it allows accurate inference of population activity, at a scale currently not accessible with experimental recordings. 

Previous works have shown how different methods can be used to model and predict noise correlations. 
The Generalised Linear Model  \cite{Pillow08} uses spike history filters to couple the spiking activity between different neurons. 
Stimulus dependent maximum entropy models \cite{Granot-Atedgi13,Ferrari18b} have coupling terms to model synchronous activity between pairs of neurons. 
These models have been successfully used to model how a population of neurons responds to a stimulus ensemble.
However, they usually fail in trying to predict the population activity in responses to different stimuli. 
This is because the parameters generating the correlations usually change when learned on different stimulus statistics \cite{Cocco09}. 
{Moreover, they were never} used to predict the activity of a population across different experiments. 
A main obstacle for this is that firing rates can vary from experiments to experiments, which would induce parameter changes in most of these models, and make generalization difficult. 
Here we have found that the noise correlation could be predicted using our model knowing just the distance between the two cells and their individual activity. 
We could then predict noise correlations across experiments. 
Having a model that generalizes across experiments is crucial to pool together recordings from neurons of the same type and reconstruct activity of large neural ensembles. 

Our ability to generalize across stimuli and even experiments demonstrates that the couplings between the recorded ganglion cells are insensitive to the context. 
This means that these couplings reflect an intrinsic property of the retinal circuit, and that the mechanism generating these noise correlations in not influenced by changes in the stimulus or in overall firing rate. Since the timescale of the observed noise correlations is very fast (few ms), the most likely mechanism is gap junctions, which create direct electrical connections between ganglion cells \cite{Bloomfield09}. 
Noise correlations might be more dependent on the context if they are generated by a shared noise source, e.g. the photoreceptor noise \cite{Ala11}. Nevertheless, our results suggest that the strength of gap junctions are tuned to a value that seems preserved between different experiments.

Our approach to model noise correlations is based on copula distributions, and asssumes that the copula parameter is constant across experiments.
An alternative, simpler approach could have been to assume that noise correlations themselves remain similar across experiments.
We constructed this simpler model by fitting an exponential function over noise correlations in our training dataset. 
This approach, however, gave significantly worse results than our method based on copulas (see supplementary sect.~\ref{suppl:nullModel}), demonstrating that our approach captures non-trivial properties of the correlated firing, i.e. its dependence on the firing rate of each cell.

Remarkably, our description of noise correlation depends only on the physical distance between the pair (see supplementary sect.~\ref{suppl:inference}). We have relaxed these assumptions and found that the copula parameter did not vary much with time, cell identity or stimulus. First, if we assume one copula parameter per time bin for a given pair of neurons, it varied little with time, and approached a constant value when the pairs' firing rate were large enough (see supplementary Fig.~\ref{figCopula}A and B). 
Second, when we inferred one parameter for each pair of cells to account for cell identities, their values still followed closely our parametric function (see supplementary Fig.~\ref{figCopula}C).
Finally, when we inferred  copula parameters from the response to different stimulus ensembles, we obtained very similar values (see supplementary Fig.~\ref{figCopula}D). 
This shows that the same copula distribution accounts well for the correlation between the two cells, independently of their firing rate, identities and of the stimulus ensemble. 

Copulas have rarely been used in neuroscience studies \cite{Berkes09,Onken09,Sacerdote12,Onken16,Wang17,Safaai18}, but none of them applied this method to predict noise correlations.
In \cite{Berkes09}, Pillow and co-workers proposed for the first time discrete copula distributions to model the total spike-count correlation in pre-motor cortex neurons.
However, they  did not distinguish stimulus from noise correlations as we have done here.

An interesting outcome of our method is the possibility to construct models of arbitrary large populations, as long as enough sequential recordings are available.
This possibility opens for testing a number of hypotheses on how correlated firing affects the overall population activity.
Previous studies \cite{Schneidman06,Tkacik14} have made conjectures on the behavior of the retinal synchronous activity at large scale by extrapolating their results from the smaller number of cells experimentally available.
The validity of these extrapolations has been recently questioned \cite{Nonnenmacher17}, pointing out how the observed correlation pattern of small systems must be different from that of larger ones. 
Here we build a large population model (Fig.~\ref{figSynth}) pooling together real data and using our validated model to infer noise correlations between cells.
Our synthetic model thus provide the framework to further test the conjectures on the system' behavior for large numbers of neurons, beyond what can be done experimentally.

Our approach is general and could be applied in any sensory area, provided that sequential recordings of the same type of cells are available, as well as some pairwise recordings to fit the model parameters. 
We applied it in the retina, where it is possible to have recordings of many neurons of the same type. 
We could thus validate our method and show that excess synchrony (i.e. beyond what could be predicted by a conditionally-independent model) increased with the number of neurons, and becomes more and more significant at larger population sizes. 
Recent technological advances will make this method relevant to understand cortical populations, where it should be soon possible to define the cell type of each recorded cell using genetic (e.g. single cell transcriptomics \cite{Economo18,Kim18}) and physiological (e.g. clustering 
of responses \cite{Baden16}) tools.
One issue could arise if noise correlations strongly depend on the stimulus, as it has, for example, been reported for V1 \cite{Banyai19}. 
In our data, noise correlations depend very little on the stimulus, which allowed us to reduce our model and {let copula parameters depend only on the distance between cells}. 
However, if noise correlations depend largely on the stimulus, the model can be extended. 
The simplest solution would be to make the copula parameters depend on the stimulus. 
If this stimulus dependence can be explicitly modeled, our method would still manage to predict the activity of large ensembles of neurons. 

Finally, here we predicted the responses of a population of neurons to a stimulus for which we have access to single cell responses across stimulus repetitions. 
If a model is available to predict the responses of single cell to other stimuli, it could be used to predict the marginal probability distribution of each individual neuron, and then combined with our approach to predict the activity of the whole population.
This makes our method complementary to recent efforts trying to model and predict accurately the response of single neurons to complex stimuli in sensory areas \cite{Mcintosh16,Cadena18,Yamins16}.

\clearpage
\section*{Methods}
\paragraph{Multi-electrode array recordings.}~
We analyze the response of rat RGCs to visual stimulation recorded in 9 multi-electrode array \textit{ex-vivo} experiments \cite{Marre12}, and spike sorted with \textit{SpyKING CIRCUS} \cite{Yger18}. 
This dataset and the experimental methods have been already previously described \cite{Deny17}.
In one experiment we probed the retinal response to three different visual stimuli: (i) a random black and white checkerboard, with spatio-temporal uncorrelated checkers; (ii) a full-field stimulus with fluctuating luminance and (iii) two gray horizontal bars performing an independent Brownian motion along the vertical direction \cite{Deny17}.
In the other 8 experimental sessions, only the response to random black and white checkerboard and full-field was retained and analyzed here.
Each of these stimulations lasted about $10 sec$ and have been repeated at least $R=79$ times.
Spiking times have been binned with a window of about $17ms$,  corresponding to a bin rate of $60Hz$.
With a custom algorithm \cite{Deny17} -similar to that of \cite{Baden16}- we used the cell's response to the full-field stimulation to identify the type of  the recorded RGCs.
Across the 9 experiments, we identified populations of $20 \pm 6$ (mean $\pm$ s.d.) OFF-Alpha cells.

\paragraph{Stimulus and noise correlations.}~ 
After binning the spiking response of each cell, we estimate $n_i^{(t,r)}$, the number of emitted spikes by cell $i$, in the time-bin $t$, during repetition $r$, and its mean across repetitions $\mu_i^{(t)}$.
Then we calculate the total covariance between two neurons $(i,j)$ as follows:
\begin{equation}
    Cov_{total}(n_i, n_j) = \frac{1}{T} \sum_{t=1}^T \frac{1}{R} \sum_{r=1}^R (n_i^{(t,r)} - \mu_i)(n_j^{(t,r)} - \mu_j)
\end{equation}
Where $\mu_i = \sum_{t=1}^T \mu_i^{(t)}/T  $ is the mean number of spikes across repetitions, and  then averaged in time.
It is possible to decompose the total covariance into a sum of the so called ``stimulus"
and ``noise" covariances.
We calculated these quantities as follows
\begin{equation}
    Cov_{noise}(n_i,n_j) = \frac{1}{T} \sum_{t=1}^T \frac{1}{R} \sum_{r=1}^R (n_i^{(t,r)} - \mu_i^{(t)})(n_j^{(t,r)} - \mu_j^{(t)})
\end{equation}
\begin{equation}
    Cov_{stimulus}(n_i,n_j) = \frac{1}{T} \sum_{t=1}^T \frac{1}{R} \sum_{r=1}^R (\mu_i^{(t)} - \mu_i)(\mu_j^{(t)} - \mu_j)
\end{equation}
Noise correlations are then estimated as:
\begin{equation}
    Corr_{noise}(n_i,n_j) = \frac{Cov_{noise}(n_i,n_j)}{\sqrt{ V_i V_j }} 
\end{equation}
where $V_i = Cov_{total}(n_i, n_i)$.

Triplet noise correlations are instead defined as:
\begin{equation}
    Corr_{noise}(n_i,n_j,n_k) = \frac{ \frac{1}{T} \sum_{t=1}^T \frac{1}{R} \sum_{r=1}^R (n_i^{(t)} - \mu_i^{(t)})(n_j^{(t)} - \mu_j^{(t)})(n_k^{(t)} - \mu_k^{(t)}) }{ \sqrt{ V_i V_j V_k }}
\end{equation}

\paragraph{Copulas.}~
Copula-based modeling allows for disentangling the marginal distributions of two random variables from their mutual dependency, that can therefore be modeled alone, without the additional difficulties due to potentially complicated marginal distributions.
Consider two random variables $X$ and $Y$, with joint distribution $f_{X,Y}$, marginal distributions $f_X$ and $f_Y$ and marginal cumulative density function (c.d.f.) $F_X$ and $F_Y$ respectively.
By construction, the random variables $U_X \equiv F_X(X), X \sim f_X $ and $U_Y \equiv F_Y(Y),Y \sim f_Y$ have uniform distributions over $[0,1]$.
Consequently the joint distribution of $(U_X,U_Y)$ has uniform marginals, yet it contains all the information about the mutual dependence between $X$ and $Y$.
This property allows us to model the dependency of $U_X$ and $U_Y$ \textit{instead} of that of $X$ and $Y$.
Specifically, a copula is the c.d.f. of the joint variable $(U_X,U_V)$,
i.e. a function $C(\cdot,\cdot): [0,1]^2 \to [0, 1]$.
and it can be used to reconstruct the joint distribution of $(X,Y)$, via its c.d.f.:
\begin{equation}
F_{X,Y}(x,y) = C\left( F_X(x) , F_Y(y) \right).
\label{eq_copula}
\end{equation}
Sklar's theorem (see supplementary sect. \ref{suppl:math}) ensures the existence and uniqueness of $C$, and this allows for modeling the mutual dependency between $X$ and $Y$, independently from their marginal distributions.

In most practical situations, the copula $C$ is chosen from a parametric copula family. 
In this work we chose to work with Frank copulas:
\begin{equation}
   C_\text{Frank}(u,v|\theta) = -\theta^{-1}\log \big( 1 + \frac{(e^{-\theta u} - 1)(e^{-\theta v} - 1)}{e^{-\theta} - 1} \big),
\label{eqFrank}
\end{equation}
where $\theta \in \mathcal{R}$ is the copula parameter, that can be estimated by log-likelihood maximization.
We chose the Frank copula because this family has already been showed to perform well in modeling spike counts \cite{Berkes09}.
Once $\th$ has been inferred, the marginal distributions can in turn be approximated either with some model, or, as we will do in this work, empirically.
We refer to the Mathematics section, the literature and textbooks (see, for example, \cite{Trivedi07}) for more explanations and details on copula models and/or other copula families.

\paragraph{Copula-based model.}~
We constructed a copula model able to predict $P(n_i^{(t)},n_j^{(t)})$, the joint distribution of pair of spike counts in time:
\begin{equation}
P\left( \{n_i^{(t)},n_j^{(t)}\}_{t=1}^T \right) = \prod_{t=1}^T c_\text{Frank}\left( F^{(t)}_i(n_i^{(t)}) , F^{(t)}_j(n_j^{(t)})~\Big|~{\hat \th}(d_{ij})~ \right)~,
\label{eq_copula_time}
\end{equation}
where $c_\text{Frank}$ is the copula density function, corresponding to the c.d.f. of Eq.~(\ref{eqFrank}), and  $F^{(t)}_i(n_i^{(t)})$ is the \textit{empirical} c.d.f. $n_i^{(t)}$, that we estimate across repetitions,  $d_{ij}$ is the distance between neurons $i$ and $j$, and ${\hat \th}(d_{ij})$ is a parametric function: ${\hat \th}(d) = \exp( a+bd+cd^2)$ and ${\hat \th}(d) = 0$ if $d>1mm $.
To infer this function from data, we first inferred a copula parameter for each pair of neurons, $\th_{ij}$, by log-likelihood maximization, and then we obtained $a=3.2$,$b=-0.013 \mu m^{-1}$ and $c = 7\, 10^{-6} \mu m^{-2}$ by fitting the behavior of $\th_{ij}$ with respect to the distance $d_{ij}$. 
See supplementary sect.~\ref{suppl:inference} for further information.

\paragraph{Time-dependent Maximum Entropy model.}~
The time-dependent maximum entropy model we used is:
\begin{equation}
P\left(  \big\{ \{ n_i^{(t)}\})_{i=1}^N \big\}_{t=1}^T \right) = \prod_{t=1}^T \left( \exp \left\{ \sum_i h_i^{(t)} n_i^{(t)} + \sum_{i\leq j} n_i^{(t)} J_{ij} n_j^{(t)} -\sum_i \ln(n_i^{(t)}!) \right\} \Big/ Z^{(t)}\right)
\label{MEmodel}
\end{equation}
where $n_i^{(t)} \in[0,1,\dots,n^\text{Max}]$ is an integer spike-count, with $n^\text{Max}$ matched from data.
The index ``$^{(t)}$'' expresses the time dependence.
$Z^{(t)}$ is a normalization constant (the partition function).
$ h_i^{(t)}$ is the local field for neuron $i$ at time $t$ imposing the firing probability and $J_{ij}$ is the couplings network that allows for reproducing the system's correlations.
Note that $J_{ij}$ does not depend on time and it includes also the diagonal terms $J_{ii}$ which set each neurons variance equal to its empirical value \cite{Ferrari18a,Ferrari18b}.
The log-factorial term allows for matching the single neurons statistics \cite{Ferrari18a}, as by taking $J=0$ the model reduces to a collection of independent Poisson distributions.

The inference of the model (\ref{MEmodel}) is done by log-likelihood maximization using an iterative algorithm with adaptive learning rate similar to that of \cite{Ferrari16}.
Thanks to its exponential form, the model inference requires only the average value of $n_i^{(t)}$ across repetitions and the value of the noise covariances (see Methods).
In our case we estimate the first from the marginal response of each neurons and we used the copula model to predict the second.

\paragraph{Synthetic lattice.}~
In order to construct a synthetic lattice that respects the inter-cells distance of empirical recordings, we started by a triangular regular lattice with side $194\mu m$, and then we added a Gaussian noise with a standard deviation of $22\mu m$ to the x and y coordinate of each cells.
These parameters are optimized in order to match the distribution of cells distances measured in real experimental recordings.

\paragraph{Population activity variance and synchrony.}~
In this paper we compute the population activity as the sum of all spike-counts, $\sum_i n_i(t)$.
To compute the ``population activity variance'' of Fig.~\ref{figSynth}C we first estimate the variance of the population activity for each time-bin.
Then we averaged over time and finally we normalized by the population size.
``Synchrony'' of Fig.~\ref{figSynth}D is instead the probability of observing an event with a population activity larger than the mean plus one standard deviation of the population activity of the shuffled model. 
We compute this for each time-bin and then we averaged over time.

\clearpage
\renewcommand{\thesection}{S\arabic{section}} 
\renewcommand{\thefigure}{S\arabic{figure}}

\setcounter{section}{0}
\setcounter{figure}{0}
\setcounter{paragraph}{0}
\section{Supplementary mathematics}
\label{suppl:math}
\paragraph{Copula's definition.}~ 
A copula function is the c.d.f. of a bi-variate distribution with $Uniform(0,1)$ marginals.
This mathematical construct allows to model the dependency structure of bi-variate random variables
separately from their marginal distributions, as the following theorem shows.

\paragraph{Sklar's theorem.}~
Let $X$ and $Y$ be any two, mutually dependent, real random variables.
Let $F_X$, $F_Y$, and $F_{(X,Y)}$ be the c.d.f.s of $X$, $Y$, and $(X,Y)$ respectively.
Note that for any $X$ we have $F_X(X) \equiv Uniform(0,1)$, \textit{idem} for $Y$.
The Sklar theorem asserts that given such $X$, and $Y$:

\begin{equation}
    \exists! \,\text{a copula } C, \text{such that } F_{X,Y}(x,y) = C(F_X(x),F_Y(y))
\end{equation}

\paragraph{Discrete copulas.}~
The proof of existence of Sklar's theorem holds for both continuous, and discrete random variables (as is our case).
However, in practice, to apply copula models in the discrete case we need to make adjustments.
In particular we turn a continuous copula, into a discrete distribution that may take values in a countable set of points in $[0,1]$.
By doing so we define a so called ``pseudo density" function, that describes a discrete counterpart to copulas.
Assuming that $(X,Y) \in \mathcal{N}^2$, without loss of generality,
the ``pseudo density" of a copula defined by a copula density $c$ is given by:

\begin{equation}
    f_{pseudo}(F_X(x),F_Y(y)) = \int_{F_X(x-1)}^{F_X(x)}\int_{F_Y(y-1)}^{F_Y(y)}c(u,v) \,du\,dv
\end{equation}

\section{Supplementary information: model construction}
\label{suppl:inference}
Here we fully justify how we simplified the copula model, from its bare version with one parameter for each neuron pair in each time-bin, to the final version with just three parameters in total.

\begin{figure}[ht!]
\centering
\includegraphics[clip=true,keepaspectratio,angle=-0,width=1.0\columnwidth]{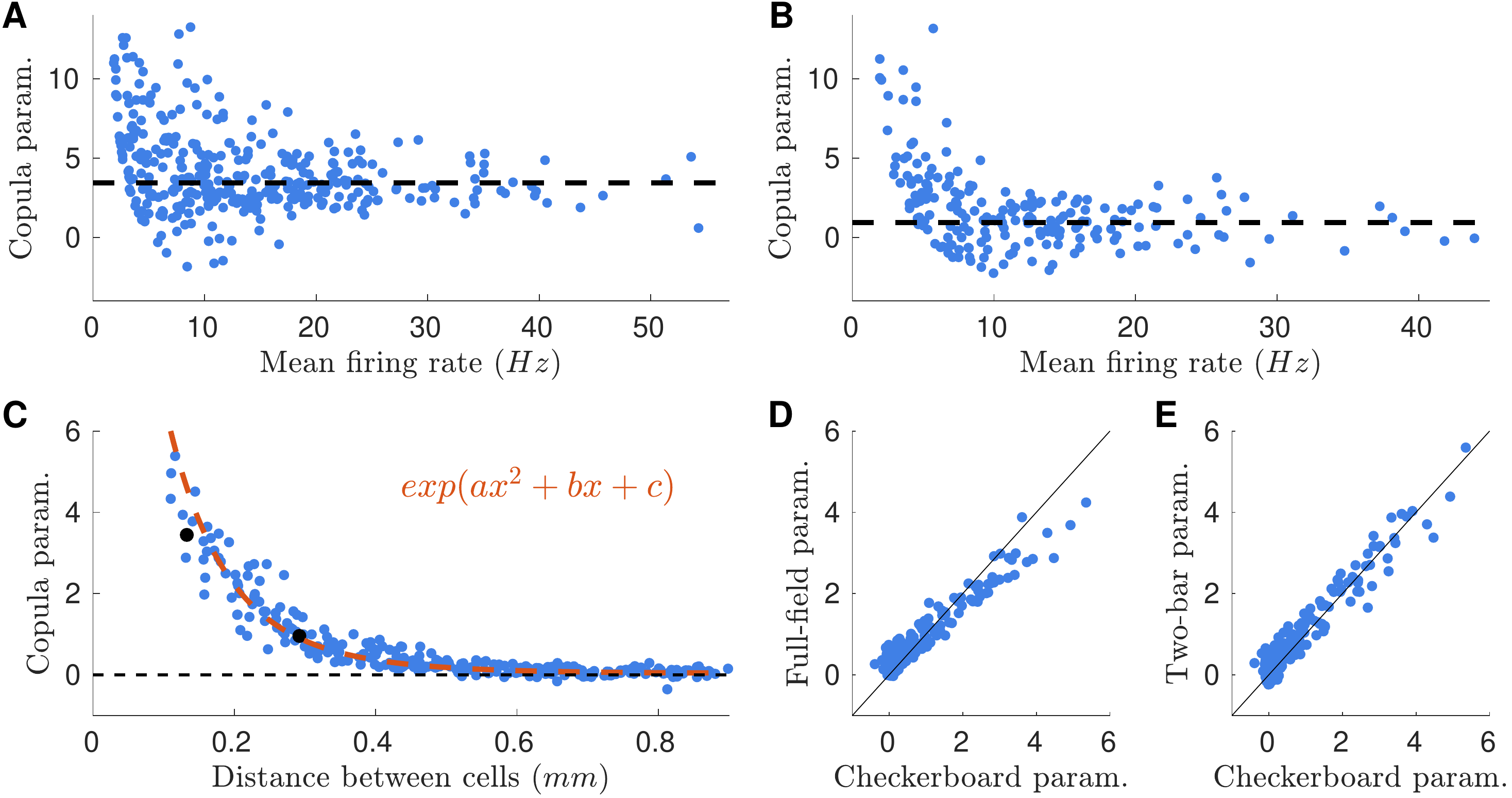}
\caption{
\textbf{Stimulus-conditioned copula model is robust across different stimulus ensembles.}
{\bf A}) Values of the inferred copula parameter in all time-bins for an example neuron pair, plotted against the mean firing rate of the two cells. 
Horizontal line correspond to the inferred parameter in the \textit{time-independent} copula model (see text)
{\bf B}) As {\bf A}, but for another example pair.
{\bf C}) Inferred Frank copula's parameters plotted against the distance between cells for checkerboard stimulation. Parameters of each pairs are inferred independently.
Orange: exponential fit used for estimating the copula parameter from the cell distance in the final version of the model.
{\bf D}) Scatterplot of the \textit{time-independent} (see text) parameters inferred from checkerboard and full-field stimulations
{\bf E}) Scatterplot of the parameters inferred from checkerboard and two-bar stimulations
}
\label{figCopula}
\end{figure}

As explained in the Methods section, our starting point is a copula model where for each neuron pair $(i,j)$ and each time-bin $t$ a copula parameter $\th_{ij}^{(t)}$ accounts for the correlation between the two spike-counts.
Despite being very accurate, this model has little capacities to generalize across stimulus conditions or experiments.
Moreover, because of the large number of parameters is potentially pruned to overfitting.
In Fig.~\ref{figCopula}A,  for an example neuron pair, we show the values of the inferred parameters in all time-bins plotted against the mean firing rate of the two cells -in the corresponding time-bin and computed across repetitions.
As can be observed, at high firing rate, that is when the statistics is large and the  inference error is small, the inferred parameters tend to accumulate around a single value.
This result suggested us that a model where the copula parameters does not depend on time could have a similar performance, yet requiring much less parameters.

In order to infer such time-independent copula parameters, for each neuron pair, we first select the ``active'' time-bins where the two neurons spiked synchronously in at least one repetition:
$\sum_{r=1}^R n_{i,r}^t  n_{j,r}^t  > 0$,
where  $n_{i,r}^t$ is the number of spikes emitted by neuron $i \in [1,\dots,N]$, in time bin $t\in [1,\dots,T]$ during repetition $r \in [1,\dots,R]$.
Once the inactive time bins are filtered out,
we estimate the model parameter by maximizing the likelihood:

\begin{equation}
    l(\theta|X) = \sum_t \sum_{r=1}^{R} \log(f_{pseudo}^{(t)}(n_{i,r}^{(t)},n_{j,r}^{(t)}|\theta))\\
                = \sum_{t} \sum_{r=1}^{R}
                    \log
                    \int_{F_i^{(t)}(n_{i,r}^{(t)} - 1)}^{F_i^{(t)}(n_{i,r}^{(t)})} 
                    \int_{F_j^{(t)}(n_{j,r}^{(t)} - 1)}^{F_j^{(t)}(n_{j,r}^{(t)})}
                    c_\theta(u,v) \,du \,dv
\label{infer_eq}
\end{equation}
where the summation over $t$ runs over the active time-bins for the neuron pair.
Note how  $F$, and thus $f_{pseudo}$, depends on $t$, as the empirical marginals are estimated separately for every time bin.
We infer the time-independent copula parameters $\th_{ij}$ by log-likelihood maximization for each neurons' pair and for checkerboard, full-field and two-bar stimuli.
Figs.~\ref{figCopula}B and C compare the parameters inferred from different stimuli and show how the inference is robust across changes of visual stimulation.
The copula parameter thus reflect some properties of the retinal network, independent of the current stimulus ensemble.

Fig.~\ref{figCopula}C show the behavior of the inferred checkerboard parameters with respect to the physical distance between cells. 
Furthermore these parameters are independent of the visual stimulus (see Figs.~\ref{figCopula}D and E).
These results suggested us that a simple fit of copula parameters may account for most of the variability of the parameter values across neuron pairs.

To further simplify our copula model, and reduce the number of its parameters, we hence fitted the inferred copula parameters with a parametric function of the inter-cell distance $d$:  $\th(d) = exp(a+b d+ c d^2)$.
The copula model takes now as input only the distance between the cells, uses it to estimate the copula parameter, and then construct the joint spike-count distribution using Eq. (\ref{eq_copula_time}).

\section{Supplementary information: simplest model for noise correlations}
\label{suppl:nullModel}
In this section we compare the performance of our copula-based approach in predicting noise correlations with a straightforward model that assumes distant-dependent noise correlations.

\begin{figure}[ht!]
\centering
\includegraphics[clip=true,keepaspectratio,angle=-0,width=1.0\columnwidth]{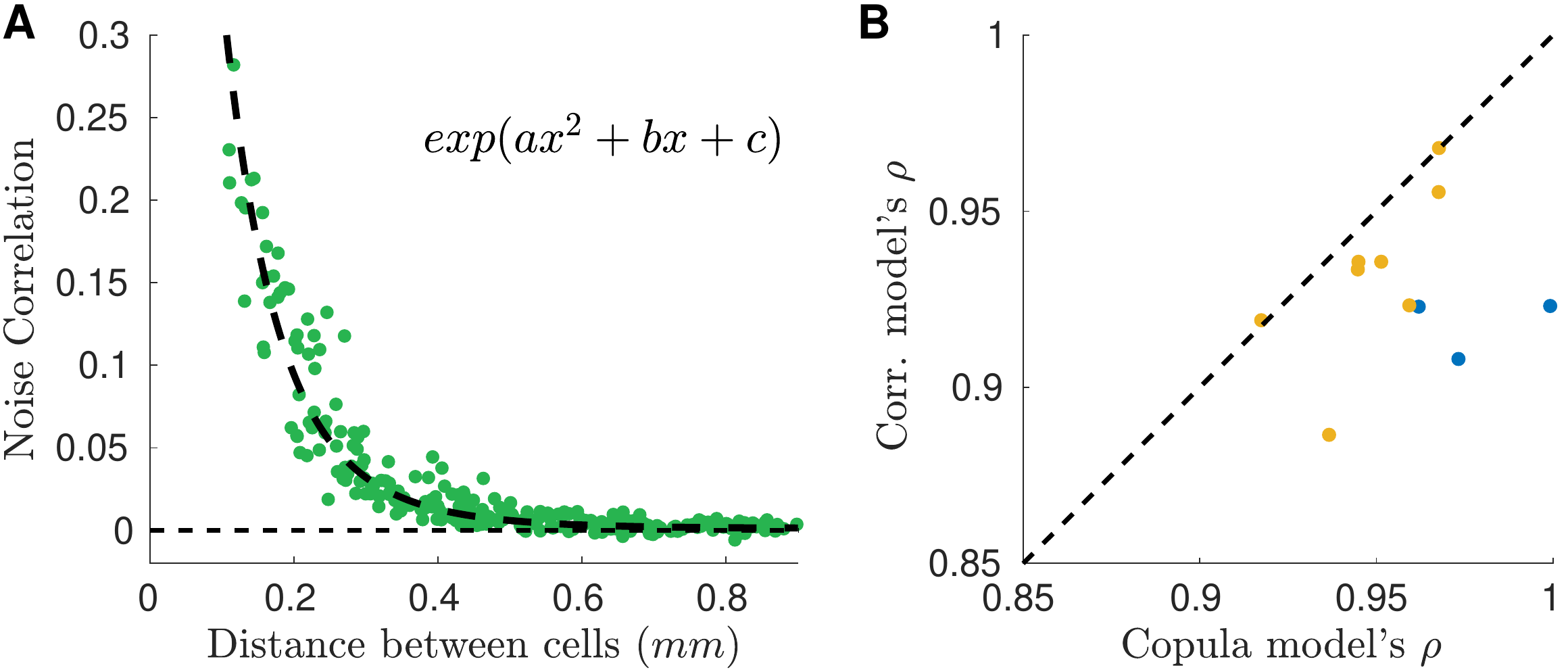}
\caption{
\textbf{Copula model outperforms simpler model with distant dependent noise-correlations.}
{\bf A}) Construction of the model: noise correlations observed in one experiment (Exp. \#1) are fitted with an exponential function of the distance between neurons. 
Such fit is then used to predict noise correlations in other experiments.
{\bf B}) Performance in predicting noise correlation for our copula model against the model presented here. Blue points: performance for the first dataset with three different stimulus ensembles.
Yellow points: performance for the other experimental sessions.
}
\label{figNull}
\end{figure}

Noise correlations decrease with the distance between the corresponding neurons (see Fig.~\ref{fig_1} and Fig.~\ref{figNull}A).
We fit this relation with an exponential function, and we asked to what extent this behavior is conserved across experiments.
To estimate this, we used this simple method to predict noise correlations in all the experiments described before.
Although the predictions were  accurate, our copula model outperforms this simpler approach (see Fig.~\ref{figNull}B).

\clearpage
\subsection*{Acknowledgments}
We like to thank M. Chalk and G. Tkacik for useful discussions.
This  work was supported by ANR Trajectory, the French State program Investissements d’Avenir managed by the Agence Nationale de la Recherche (LIFESENSES; ANR-10-LABX-65),  EC Grant  No.  H2020-785907  from  the  Human  Brain  Project, NIH Grant No. U01NS090501, and an AVIESAN-UNADEV grant to O.M. 

\nolinenumbers

\bibliographystyle{unsrt}

\end{document}